\documentclass[onecollarge,referee]{svjour2}

\usepackage{graphicx} % Include figure files
\usepackage{dcolumn}% Align table columns on decimal point
\usepackage{color}% color package
\usepackage{amsmath,amssymb}
\journalname{}
\begin{document}
\title{Pattern formation in growing sandpiles with multiple sources or sinks}
\author{Tridib Sadhu \and Deepak Dhar}
\institute{Department of Theoretical Physics, \\
Tata Institute of Fundamental Research, \\
Homi Bhaba Road, Mumbai 400005, India. \\
\email{tridib@theory.tifr.res.in; ddhar@theory.tifr.res.in}}

\date{\today}
%\pacs{89.75.Kd}{Pattern formation in complex systems}
%\pacs{45.70.Cc}{Sandpile models}
\maketitle

\begin{abstract}
Adding sand grains at a single site in Abelian sandpile models produces beautiful
but complex patterns. We study the effect of sink sites on such patterns. Sinks
change the scaling of the diameter of the pattern  with the number $N$ of sand
grains added. For example, in two dimensions, in presence of a sink site, the
diameter of the pattern grows as $\sqrt{(N/\log N)}$ for large $N$, whereas it
grows as $\sqrt{N}$ if there are no sink sites. In presence of a line of sink
sites, this rate reduces to $N^{1/3}$. We determine the growth rates for these sink geometries along with the case when there are
two lines of sink sites forming a wedge, and its generalization to higher dimensions.
We characterize one such asymptotic patterns on the two-dimensional F-lattice
with a single source adjacent to a line of sink sites, in terms of position of different
spatial features in the pattern. For this lattice, we also provide an exact
characterization of the pattern with two sources, when the line joining them is along one of the axes. 
\end{abstract}

%%%%%%%%%%%%%%%%%%%%%%%%%%%%%%%%%%%%%%%%%%%%%%%%%%%%%%%%%%%%%%%%%%%%%
\section{Introduction}
It is well known that beautiful and complex patterns can be generated by
deterministic evolution of systems under simple local rules, e.g. in the
game of life \cite{earlierone}, and Turing patterns \cite{earliertwo}.
Growing sandpiles on a flat table with boundaries by adding particles at
a constant rate gives rise to  singular structures like ridges in the
stationary state, which have attracted much attention recently \cite{hadeler,falcone_vita}.   
In the Abelian sandpile model, growing sandpiles produce richer and hence
more interesting patterns. This model is inspired by real sandpile dynamics,
but has different rules of evolution. The steady state of sandpile models
with slow driving, and presence of a boundary has been studied much in the
context of self-organized criticality \cite{dd_physica}. The Abelian sandpile
model, with particles added at one site, on an infinite lattice have the very
interesting property of {\it proportionate growth} \cite{epl}. This is
a well-known feature of biological growth in animals, where different parts
of the growing animal grow at roughly the same rate. Our interest in studying
growing sandpiles comes from  it being  the prototypical model of proportionate
growth. Most of the other growth models studied in physics literature, such
as diffusion-limited aggregation, or surface deposition do not show this
property, and the growth is confined to some active outer region, and the
inner structures once formed are frozen in, and do not evolve further in time \cite{herrmann}. 

In \cite{epl}, we studied growing sandpiles in the Abelian model on
the F-lattice and the Manhattan lattice. These are directed variants
of the square lattice, obtained by assigning directions to the bonds,
as shown in Fig.1. We found that for a particular choice of the initial
background configuration, the patterns formed can be characterized exactly.
The special  initial configuration is the one in which each alternate
site of the lattice is occupied, forming a chequerboard pattern. If
we add particles at the origin, and relax the configuration using the
sandpile toppling rules, we generate a fairly complex pattern made
up of triangles and dart-shaped patches (Fig.\ref{fig:flattice}),
that shows proportionate growth. The full characterization of this
pattern reveals an interesting underlying mathematical structure,
which seems to deserve further exploration. This is what we do in
this paper, by adding sink sites, or multiple sources.  
 
Presence of sink sites changes the pattern in interesting ways. In
particular, it changes how different spatial lengths in the pattern
scale with the number of added grains $N$. For example, in absence of
sink sites, the diameter of the pattern grows as $\sqrt{N}$ for large
$N$, whereas presence of a single sink site next to the site of addition,
this changes to a $\sqrt{N/\log{N}}$ growth. If there is a line of
sink sites next to the site of addition the growth rate is $N^{1/3}$.
We also studied the case where the source site is at the corner of a
wedge-shaped region of wedge angle $\omega=\pi/2$, $3\pi/2$, or $2\pi$,
and where the wedge boundaries are absorbing. (The last case corresponds
to the source next to an infinite half line.) For the single point source,
determination of different distances in the pattern required a solution
of the Laplace equation on a discrete Riemann surface of two sheets.
Interestingly, for these wedge angles, we still have to solve the
discrete Laplace equation, but the structure of the Riemann surface
changes, e.g. from two-sheets to three-sheets for $\omega = \pi$, and
five-sheets for $\omega = 2 \pi$.  We characterize the patterns in
terms of the solution of the discrete Laplace equation. We also show
that the pattern grows as $N^{\alpha}$, with $\alpha=2\omega/(\pi+4\omega)$.
\begin{figure}
 \begin{center}
 \includegraphics[scale=0.19,angle=90]{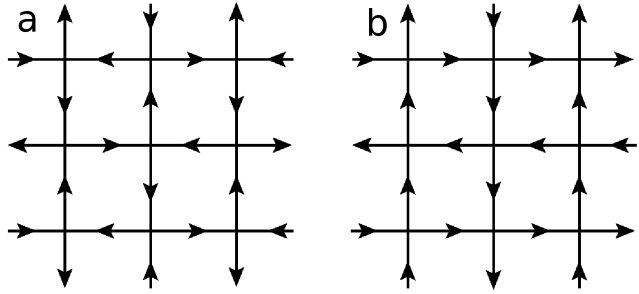}
 \caption{Directed square lattices studied in this paper:(a)  F-lattice and (b) Manhattan lattice.}
 \label{fig:lattice}
 \end{center}
\end{figure}
\begin{figure}
\begin{center}
  \includegraphics[scale=0.25]{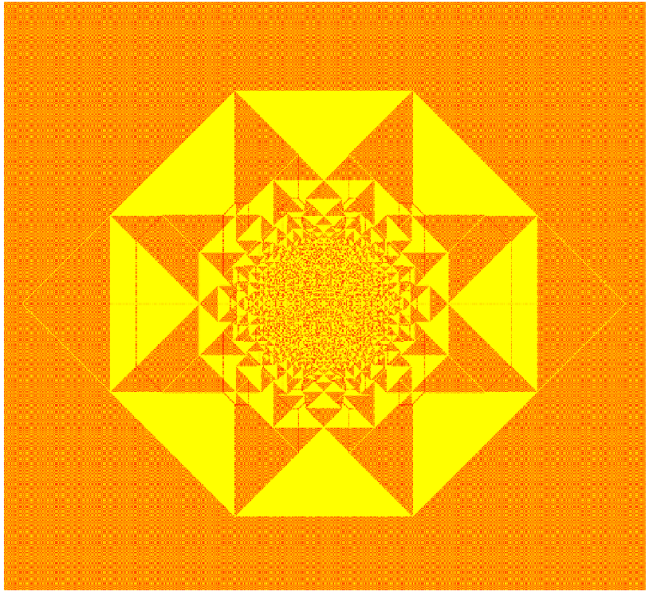}
  \caption{Stable configuration for the ASM, obtained 
  by adding $5\times10^4$ grains at one site, on the F-lattice of Fig.\ref{fig:lattice}($a$) with initial 
  chequerboard configuration. Color code: red=0, yellow=1. Apparent orange regions in the picture represent
  patches with chequerboard configuration. (Details can be seen in the online version using zoom in.)}
  \label{fig:flattice}
\end{center}
\end{figure}
 
We also study the effect of having multiple sites of addition on the pattern.
For multiple sources, the pattern of small patches near each source is not
substantially different from a single-source pattern, but some rearrangement
occurs in the larger outer patches. Two patches may sometimes join into one,
or conversely, a patch may break up into two. But the number of patches
undergoing such changes is finite. However, the sizes of all patches are
affected by the presence of other sources, and we show how these changes
can be calculated exactly for the asymptotic pattern. Spatial patterns in
sandpile models were first discussed by Liu et al \cite{liu}. The asymptotic
shape  of the boundaries of sandpile patterns produced by adding grains at 
single site on different periodic backgrounds was discussed in \cite{dhar99}.
Borgne \cite{borgne} obtained bounds on  rate of growth of these boundaries
and later these bounds are improved by Fey \textit{et al} \cite{redig} and
Levine \textit{et al} \cite{lionel}. The first detailed analysis of different
periodic  structures  found in the patterns are carried out by Ostojic  in
\cite{ostojic}. Other special configurations in the ASM models, like the
identity \cite{borgne,identity,caracciolo}, or the stable state produced
from special unstable states also show complex internal self-similar structures
\cite{liu}, which also share common features with the patterns studied here.
There are other models, which are related to the Abelian sandpile model, e.g.
the internal Diffusion-Limited Aggregation (DLA), Eulerian walkers (also called
rotor-router model), and the infinitely-divisible sandpile, which  also show
similar structure. For the Internal DLA,  Gravner and Quastel showed that the
asymptotic shape of the growth pattern is related to the classical Stefan
problem in hydrodynamics, and determined the exact radius of the pattern with
a single point source \cite{gravner}. Levine and Peres have studied patterns
with multiple sources in these models recently, and proved the existence
of a limit shape\cite{levine_peres}.

This paper is organized as follows. After defining the model in Section $2$,
we discuss scaling of the diameter of the patterns with $N$. We first consider
in Section $3$ the pattern in the presence of a line of sink sites. In Section
$4$, this analysis is extended to other sink geometries: two intersecting line
sinks in two dimensions and two or three intersecting plane sinks in three
dimensions. The case of a single sink site is a bit different from others, and
is discussed separately in Section $5$. The remaining sections are devoted to
a detailed characterization of some of these patterns. In Section $6$ we give a
summary of our earlier work on characterization of single source pattern, and
use it to characterize  the pattern in presence of a line sink. In Section $7$,
we discuss the case when there are two sources present. Section $8$ contains a
summary and  some concluding remarks.

\section{Definition of the model}
We consider the Abelian sandpile model on the F-lattice (Fig.\ref{fig:lattice}$a$).
This is a square lattice with directed bonds such that each site has two inward
and two outward arrows. A different assignment of arrow directions, that gives
us the Manhattan lattice is shown in Fig.\ref{fig:lattice}$b$. 
The asymptotic pattern formed by growing sandpile on the Manhattan lattice is the
same as on the F-lattice \cite{epl}. We shall discuss here only the F-lattice,
but the discussion is equally applicable to the Manhattan lattice.

Define a position vector on the lattice, $\mathbf{R}\equiv\left( x, y \right)$.
In the Abelian sandpile model, a height variable
$z\left( \mathbf{R} \right)$, called the number of grains on the site,  is assigned to each site $\mathbf{R}$. 
In a stable configuration all sites have height $z\left( \mathbf{R} \right)<2$.
The system is driven by adding grains at a single site and if this addition
makes the system unstable it relaxes by the toppling rule: each unstable 
site transfers one grain each in the direction of its outward arrows. 
We start with an initial configuration in which  $z\left( \mathbf{R} \right)=1$,
for sites with $(x+y)=$ even, and $0$ otherwise.
For numerical purpose we used a lattice large enough so that
none of the avalanches reaches the boundary.
The result of adding $N=5\times 10^4$ grains at the origin  is shown in Fig.\ref{fig:flattice}. 
\section{Growth of the pattern with line sink}
\begin{figure}
\begin{center}
\includegraphics[scale=0.20]{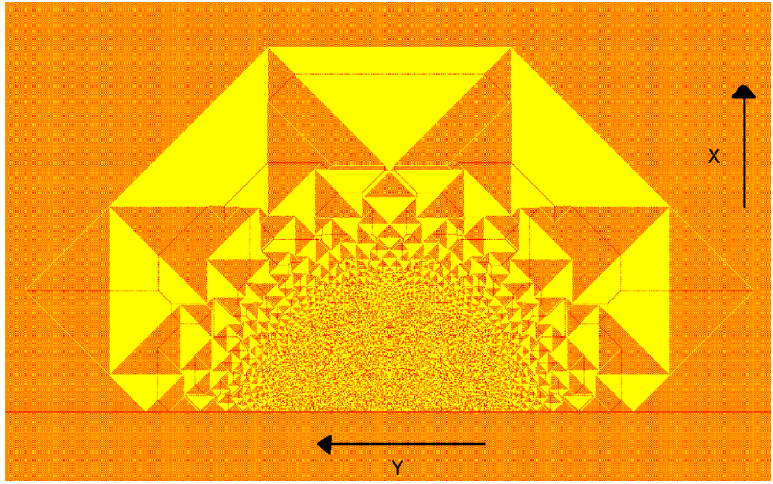}
\caption{Pattern produced by adding grains at a single site adjacent to a line
of sink sites. Color code: red=0 and yellow=1. Apparent orange regions in the picture represent
patches with chequerboard configuration.  (Zoom in for details  in the online version.)}
\label{fig:lsone}
\end{center}
\end{figure}

Consider the pattern formed by adding sand grains at a single site in
presence of a line of sink sites. Any grain reaching
a sink site gets absorbed, and is removed from the system. For simplicity let us consider the
source site at $\mathbf{R}_{o}\equiv\left(x_{o}, 0  \right)$ and the sink sites along
the $y$-axis. A picture of the  pattern produced by adding $14336000$ grains at
$\left( 1, 0 \right)$ is shown in Fig.\ref{fig:lsone}. When $N$ grains
have been added, let $2\Lambda\left( N \right)$ be the diameter of the pattern, measured
as the height of the smallest rectangle that enclosed all sites that have toppled at
least once.  We want to study how $\Lambda(N)$ increases as a function of $N$.

As mentioned before, the pattern exhibits proportionate growth. While there
is as yet no rigorous proof of this important property, we assume this in
the following. Then, it is natural to describe the pattern in reduced
coordinates defined by $\xi=x/\Lambda$ and $\eta=y/\Lambda$. A position
vector in this reduced coordinate is defined by $\mathbf{r}=\mathbf{R}/\Lambda\equiv\left( \xi, \eta \right)$.
Then in $\Lambda\rightarrow\infty$, the pattern can be characterized by a
function $\Delta\rho(\mathbf{r})$ which gives local excess density of sand
grains in the pattern in a small rectangle of size $\delta\xi\delta\eta$
about the point $\mathbf{r}$, with $1/\Lambda\ll\delta\xi$, $\delta\eta\ll1$.

Let $T_{\Lambda}\left( \mathbf{R} \right)$ be the number of toppling at site
$\mathbf{R}$ when the diameter reaches value $2\Lambda$ for the first time.
Define
\begin{equation}
\phi\left( \mathbf{r} \right)=\lim_{\Lambda\rightarrow\infty}\frac{1}{2\Lambda^{2}}T_{\Lambda}\left( \mathbf{R}' \right),
\label{phi}
\end{equation}
where $\mathbf{R}'\equiv\left(\lfloor\Lambda \xi \rfloor, \lfloor\Lambda \eta \rfloor\right)$
with $\lfloor x \rfloor$ being the floor function which gives the largest
integer $\le x$. 

From the conservation of sand-grains in the toppling process, it is easy to see that  $\phi$ satisfies the
Poisson equation \cite{epl}
\begin{equation}
\nabla^{2}\phi\left( \mathbf{r} \right)=\Delta\rho\left( \mathbf{r} \right)-\frac{N}{\Lambda^{2}}\delta\left( \mathbf{r} - \mathbf{r}_{o} \right)
\label{poisson}
\end{equation}
for all $\mathbf{r}$ in the right-half plane with $\xi > 0$,
where $\mathbf{r}_{o}$ is the position of the source in reduced
coordinates. Also as there are no toppling at sink sites, $\phi$
must satisfy the boundary condition
\begin{equation}
\phi\left( \mathbf{r} \right)=0 \rm{~ ~ for~all~} \mathbf{r}\equiv\left( 0, \eta \right).
\label{bc}
\end{equation}
A complete information of $\phi\left( \mathbf{r} \right)$ determines
the density function $\Delta\rho\left( \mathbf{r} \right)$ and in turn
characterizes the asymptotic pattern.

We can think of $\phi$ as the potential due to a point charge $N/\Lambda^{2}$ at $\mathbf{r}_{o}$
and an areal charge density $-\Delta\rho\left( \mathbf{r} \right)$, in presence
of a grounded conducting line along the $\eta$-axis. This problem can be solved
using the well-known method of images in electrostatics. Let $\mathbf{r}'$ be
the image point of $\mathbf{r}$ with respect the $\eta$-axis.
Define $\Delta\rho\left( \mathbf{r} \right)$ in the left half plane as
\begin{equation}
\Delta\rho\left( \mathbf{r}' \right)=-\Delta\rho\left( \mathbf{r} \right).
\label{imden}
\end{equation}
Then the Poisson equation for this new charge configuration is
\begin{equation}
\nabla^2\phi(\mathbf{r})=\Delta\rho(\mathbf{r})-\frac{N}{\Lambda^2}\delta\left( \mathbf{r} - \mathbf{r}_{o} \right)+\frac{N}{\Lambda^2}\delta\left( \mathbf{r} - \mathbf{r'}_{o} \right).
\label{poisson4}
\end{equation}
As the function $\Delta\rho\left( \mathbf{r} \right)$ is odd under reflection,
$\phi$ automatically vanishes along the $\eta$-axis.

We define $N_{r}$ as the number of sand grains that remain unabsorbed. Then
\begin{equation}
N_{r}=\sum_{x>0}\sum_{y}\Delta z\left( x, y \right),
\label{nr1}
\end{equation}
where $\Delta z\left(x, y  \right)$ is the change in height variables
before and after the system relaxes. Clearly, for large $\Lambda$, we can
write
\begin{equation}
N_{r}\simeq\Lambda^{2}\int_{\mathbb{H}}d\tau\Delta\rho\left( \mathbf{r} \right),
\label{nr2}
\end{equation}
where $d\tau=d\xi d\eta$ is the infinitesimal area around $\mathbf{r}\equiv \left( \xi, \eta \right)$ and
the integration performed over the right half-plane $\mathbb{H}$ with $\xi>0$.
We shall use the sign $\simeq$ to denote equality up to leading order in $\Lambda$. 
Since $\Delta\rho\left( \mathbf{r} \right)$ is a non-negative bounded function,
exactly zero outside a finite region, this integral exists. Let its value be $C_{2}$ and then we have
\begin{equation}
N_{r}\simeq C_{2}\Lambda^{2}.
\label{nr3}
\end{equation}

Let $N_{a}$ denote the number of grains that are absorbed by the sink
sites. Then considering that grains can reach sink sites only by toppling at its neighbors we have
\begin{equation}
N_{a}\simeq\frac{1}{2}\sum_{y}T_{\Lambda}\left( 1, y \right).
\label{na1}
\end{equation}
The factor $1/2$ comes from the fact that in F-lattice, only half of the sites
on the column $x=1$ would have arrows going out to the sink sites. Then using
our scaling ansatz in equation (\ref{phi}), for $\Lambda$ large,
\begin{equation}
T_\Lambda\left( 1, y \right)\simeq 2 \Lambda  \left.\frac{\partial \phi}{\partial \xi}\right \arrowvert_{\xi=0}.
\label{T1}
\end{equation}
Hence
\begin{equation}
N_{a}\simeq\Lambda^{2}\int_{-\infty}^{\infty}d\eta\left.\frac{\partial \phi}{\partial \xi}\right \arrowvert_{\xi=0}.
\label{na2}
\end{equation}
Now from equation (\ref{poisson4}) the potential $\phi$ can be written as sum of two
terms: $\phi_{dipole}$ due to two point charges $N/\Lambda^{2}$ and $-N/\Lambda^{2}$
at $\mathbf{r}_{o}\equiv\left( \xi_{o}, 0 \right)$ and its image point
$\mathbf{r}'_{o}\equiv\left( -\xi_{o}, 0 \right)$ respectively, and the term 
$\phi_{rest}$ due to the areal charge density.
\begin{equation}
\phi\left( \mathbf{r} \right)=\phi_{dipole}\left( \mathbf{r} \right)+\phi_{rest}\left( \mathbf{r} \right),
\label{phi2}
\end{equation}
where
\begin{eqnarray}
\nabla^{2}\phi_{dipole}\left( \mathbf{r} \right)&=&-\frac{N}{\Lambda^2}\delta\left( \mathbf{r} - \mathbf{r}_{o} \right) + \frac{N}{\Lambda^2}\delta\left( \mathbf{r} - \mathbf{r}'_{o} \right),\nonumber \\
\nabla^{2}\phi_{rest}\left( \mathbf{r} \right)&=&\Delta\rho\left( \mathbf{r} \right).
\label{poisson5}
\end{eqnarray}
We first consider the case where $R_{o}$ is finite and $r_{o}=R_{o}/\Lambda$ vanishes
in the large $\Lambda$ limit. Then $\phi_{dipole}$ reduces to a dipole potential,
and it diverges near origin. However,  $\phi_{rest}\left( \mathbf{r} \right)$ is a
non-singular function for all $\mathbf{r}$. From the solution of
dipole potential, it is easy to show that
\begin{equation}
\phi_{dipole}\left( r, \theta \right) \approx  A \frac{\cos\theta}{r},
\label{dipole1}
\end{equation}
for $1\gg r\gg 1/\Lambda$, where we have used polar coordinates $\left( r, \theta \right)$
with $\theta$ being measured with respect to the $\xi$-axis. Here 
$A$ is a numerical  constant, which is determined by the property of the asymptotic pattern. Then
\begin{equation}
\left.\frac{\partial \phi}{\partial \xi}\right \arrowvert_{\xi=0}=\frac{A}{\eta^{2}},
\label{phi3}
\end{equation}
and the integral in equation (\ref{na2}) diverges as $A/\eta_{min}$, where
$\eta_{min}$ is the cutoff introduced by the lattice. Using $\eta_{min}=\mathcal{O}\left( 1/\Lambda \right)$
it is easy to show that
\begin{equation}
N_{a}\simeq C_{1}\Lambda^{3},
\label{na3}
\end{equation}
where $C_{1}$ is a constant. Then using equation (\ref{nr3}) and (\ref{na3}) and that $N_{a}$ and $N_{r}$ add up to $N$,
we get
\begin{equation}
C_{1}\Lambda^{3}+C_{2}\Lambda^{2}\doteqdot N.
\label{scalethree}
\end{equation}
Here we use the symbol $\doteqdot$ to denote ``nearly equal to''.
Considering the dominant term in the expression for large $\Lambda$, it follows
that $\Lambda$ increases as $N^{1/3}$.

The above scaling behavior is verified with our numerical data.
Let $\Lambda^{\ast}\left( N \right)$ be the real positive root of the
equation (\ref{scalethree}) for a given value of integer $N$. As
$\Lambda$ takes only integer values on the lattice, an estimate of
it would be $Nint\left[ \Lambda^{\ast}(N) \right]$ the integer nearest
to $\Lambda^{\ast}\left( N \right)$.

Interestingly, we found that for
a choice of $C_{1}=0.1853$ and $C_{2}=0.528$, this estimate gives
values which differ from the measured value $\Lambda\left( N \right)$
at most by $1$ for all $N$ in the range of $100$ to $3\times10^{6}$.
Clearly more precise estimates of $C_{1}$ and $C_{2}$ would be
required if we want this to work for larger $N$. Here we find
Eqs.(\ref{nr3}) and (\ref{na3}) on dimensional counting grounds, and
the final Eq.(\ref{scalethree}) is then only a statement
of conservation of sand grains. It is quite remarkable that this scaling
analysis gives almost the exact value of $\Lambda\left( N \right)$.
The equation has an important feature. It includes ``correction to
scaling'' term whereas the usual scaling analysis ignores the sub-leading powers.

For patterns in the other limit where the source is placed at a distance
$\mathcal{O}\left( \Lambda \right)$ such that $r_{o}$ is non zero for $\Lambda\rightarrow\infty$,
$\phi_{dipole}$ is non-singular along  the  sink line. Then, clearly  $N_{a}\sim \Lambda^{2}$ and as a result $\Lambda\left( N \right)\sim N^{1/2}$.

\section{Generalization to more complex patterns}
The above analysis can be easily generalized to a case
with sink sites along two straight lines intersecting at an angle $\omega$
and a point source inside the wedge. For square lattice,
$\omega=0, \pi/2, \pi, 3\pi/2$ and $2\pi$ are most easily constructed,
and avoid  problems of lines with irrational slopes, or
slopes of rational numbers with large denominators. The wedge with wedge-angle 
$\omega=\pi/2$ is obtained by placing sink sites along the $x$ and $y$-axis
and the source site at $\mathbf{R}_{o}\equiv\left( 1, 1 \right)$ in
the first quadrant. The pattern with line sink, discussed in
previous section, correspond to $\omega=\pi$.

For any general $\omega$, corresponding electrostatic problem reduces to
determining the potential function $\phi$ inside a wedge formed by two
intersecting grounded conducting lines. Again the potential has two
contributions: the potential $\phi_{point}\left(\mathbf{r}  \right)$
due to a point charge at the source site and the potential $\phi_{rest}\left( \mathbf{r} \right)$
due to the areal charge density. We first consider the case where the source site is placed at a finite
distance from the wedge corner such that the distance in reduced
coordinate vanishes for $\Lambda$ large limit. In this limit $\phi_{rest}$ is non-singular
function of $\mathbf{r}$ while $\phi_{point}$ diverges close
to the origin. A simple  calculation of the electrostatic problem gives
\begin{equation}
\phi_{point}\left( r, \theta \right)= A \frac{\sin \alpha \theta}{r^{\alpha}},
\label{mpole}
\end{equation}
where $\alpha=\pi/\omega$ and we have used polar coordinates $\left( r, \theta \right)$ with
the polar angle $\theta$ measured from one of the absorbing lines.
Again $A$ is a constant independent of $N$ or $\Lambda$ and is a property of the asymptotic pattern.
Then arguing as before, we get
\begin{equation}
N_{a}\simeq C_{1}\Lambda^{2+\alpha}\rm{~ ~ and ~ ~ }N_{r}\simeq C_{2}\Lambda^{2}.
\label{na4}
\end{equation}
So the equation analogous to equation (\ref{scalethree}) is
\begin{equation}
C_{1}\Lambda^{2+\alpha}+C_{2}\Lambda^{2}\doteqdot N.
\label{scalefour}
\end{equation}
For wedge angle $\omega=\pi$, $\alpha = 1$, and the above equation reduces to 
Eq.(\ref{scalethree}). For $\omega = 2\pi$, the value of $\alpha$ is $1/2$.
Again, Eq. (\ref{scalefour}) is in very good agreement with our numerical data.
Let $\Lambda^{\ast}(N)$ be the solution of Eq. (\ref{scalefour}) for a given $N$.
Choosing $C_{1}=0.863408$ and $C_{2}=0.043311$, we find that the function
$Nint\left[ \Lambda^{\ast}\left( N \right) \right]$
differ from the measured values of $\Lambda$ at most by $1$ for all $N$ in the range from $100$ to $2\times10^{5}$.

For the problem where the source site is at a distance
$\mathcal{O}\left( \Lambda \right)$ from the wedge corner 
both the functions $\phi_{rest}$ and $\phi_{point}$ are nonsingular
close to the origin.
Then it is easy to show that $\Lambda\left( N \right)$ grows as $N^{1/2}$.

The  argument is easily  extended to other lattices with
different initial height distributions, or to higher dimensions.
Consider, for example, an Abelian sandpile model defined on the  cubic 
lattice.  Allowed heights are $0$ to $5$, and a site topples if the height
exceeds $5$, and sends one particle to each neighbor. The sites are labelled
by the Cartesian coordinates $(x,y,z)$, where $x,y$ and $z$ are integers. We
consider the infinite octant defined by $x\geq 0,y \geq 0,z \geq 0$. We start
with all heights $4$, and add sand grains at the site $(1,1,1)$.  We assume
that the sites on planes $x=0$, $y=0$ and $z=0$ are all sink sites, and any grain reaching there is lost.
We add $N$ grains and determine the diameter of the resulting stable pattern. 

We again reduce the potential function in two parts: $\phi_{point}$ due to a point charge at 
$\left( 1/\Lambda,1/\Lambda,1/\Lambda \right)$ and $\phi_{rest}$ due to bulk charge density in presence
of three conducting grounded planes. Then simple electrostatic
calculation gives that the potential $\phi_{point}$ is the octapolar
potential of the form
\begin{equation}
\phi\left( r, \theta, \Phi \right)\simeq \frac{f\left( \theta,\phi \right)}{r^{4}},
\label{octupole}
\end{equation}
where the spherical coordinate is used to denote position.
This then implies that the equation determining the dependence
of $\Lambda$ on $N$ is
\begin{equation}
C_{1}\Lambda^{6}+C_{2}\Lambda^{3}\doteqdot N
\label{scalefive}
\end{equation}
Like the other cases, this relation is also confirmed against numerical data.
The function $Nint\left[ \Lambda^{\ast}\left( N \right) \right]$ with $C_{1}=0.0159$
and $C_{2}=88$ gives almost exact values of $\Lambda\left( N \right)$. We have checked
for $N$ between $5\times10^{5}$ to $5\times10^{8}$, the  difference is at most  $1$.
\section{A single sink site}
\begin{figure}
\begin{center}
\includegraphics[scale=0.24]{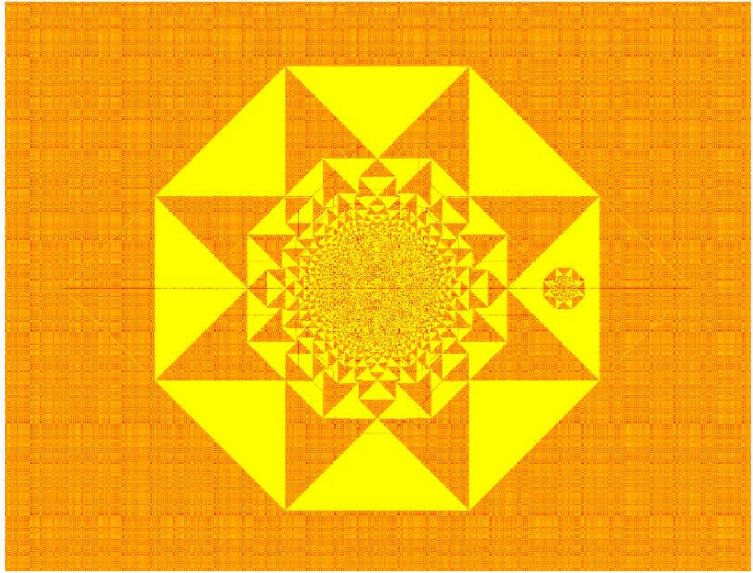}
\end{center}
\caption{Pattern produced by adding $224000$ grains at the origin
with a sink site at $\left( 400,0 \right)$ inside a patch of density $1$ (color yellow ). Color code red$=0$ and
yellow$=1$. The apparent orange regions correspond to chequerboard height distribution. (Zoom in for details in the online version.)}
\label{fig:psone}
\end{figure}

Let the site of addition is the origin and the sink site is placed at $\mathbf{R}_{o}$.
We will show that when  $\mathbf{r}_{o}$ lies in a dense patch (color yellow in Fig.\ref{fig:psone}), the asymptotic patterns are identical to the one
produced in absence of a sink.

The patterns produced for $r_{o}$ close to $1$ with sink sites placed deep inside 
a dense patch is simple to analyze, even for finite but large $\Lambda$. One such pattern
is presented in Fig.\ref{fig:psone}. 
\begin{figure}
\begin{center}
\includegraphics[scale=0.25]{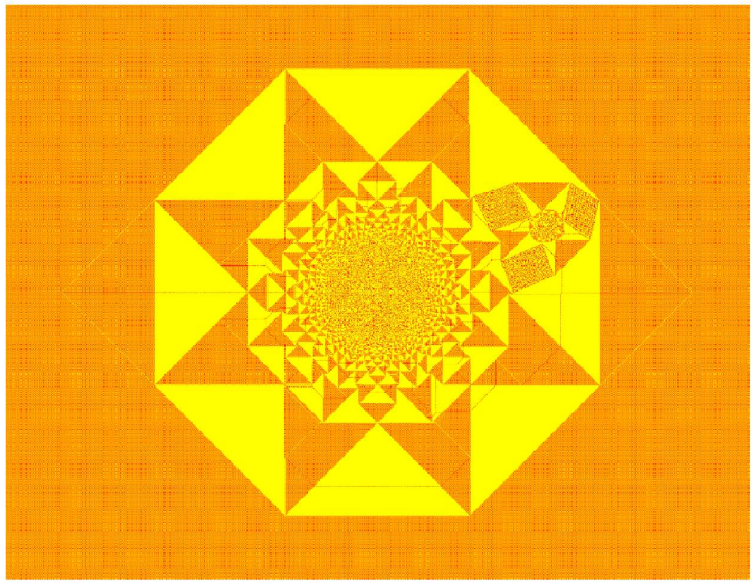}
\end{center}
\caption{Pattern produced by adding $224000$ grains at origin
with a sink site placed at $\left( 360,140 \right)$ inside a low density patch. 
Color code red$=0$ and
yellow$=1$. The apparent orange regions correspond to the chequerboard height distribution. (Details can be seen in the online version using zoom in.)}
\label{fig:pstwo}
\end{figure}

We see that the effect of sink site on the pattern is to produce a
depletion pattern centered at the sink site. The depletion pattern
is a smaller copy of the single source pattern. We define the function
$\Delta z_{sink}\left( \mathbf{R}; N \right)$ as the difference between
the heights at ${\mathbf R}$ in the final stable configuration produced
by adding $N$ grains at the origin, with and without sink. 
\begin{equation}
\Delta z_{sink} \left( \mathbf{R}; N \right)= \Delta z_{source+sink}\left( {\mathbf R}; N \right)- \Delta z_{source}\left( \mathbf{R}; N \right). 
\end{equation}
From the figure it is seen that, in this case, $\Delta z_{sink}\left( \mathbf{R}; N \right)$
is negative of the pattern produced by a smaller source, centered at $\mathbf{R}_0$.
The number of particles required to produce this smaller pattern is exactly the number of particles
$N_a$ absorbed at the sink site.
\begin{equation}
\Delta z_{sink}\left( \mathbf{R}; N \right)= - \Delta z_{source}\left( \mathbf{R}-\mathbf{R}_{o}; N_{a} \right).
\label{zsink}
\end{equation}
This is immediately seen from the fact that the 
toppling function $T_\Lambda\left( \mathbf{R} \right)$ satisfies 
\begin{equation}
\sum_{\mathbf{R'}\in\Gamma(\mathbf{R})} T_{\Lambda}\left( \mathbf{R'} \right)-2T_{\Lambda}\left( \mathbf{R} \right)=\Delta z_{source+sink}\left( \mathbf{R}; N \right)-N \delta_{\mathbf{R}, \mathbf{0}} + N_{a}\delta_{\mathbf{R}, \mathbf{R}_{o}},
\label{discrpois}
\end{equation}
where  $\Gamma(\mathbf{R})$ is the set of two neighbors
which transfer grains to the site $\mathbf{R}$ under toppling.  
Let $T_{source} ({\mathbf R}; N)$ be the number of toppling at ${\mathbf R}$, when we add $N$ particles at the origin. Since Eq. (\ref{discrpois}) is  a linear equation, it follows that a solution of this equation is
\begin{equation}
T_{\Lambda}\left( \mathbf{R} \right)=T_{source}\left( \mathbf{R}; N \right)-T_{source}\left( \mathbf{R}-\mathbf{R}_{o}; N_{a} \right),
\label{lcomb}
\end{equation}
This is a valid solution of our problem, if the corresponding heights in the final configuration with sink are all non-negative.
This happens when the region with nonzero $\Delta z_{sink}$ is confined within the dense patch.

The number $N_{a}$  can be determined from the requirement that number of toppling at the sink site is zero.  
The potential function for a single source problem diverges as $\left( 4\pi \right)^{-1}\log r$
near the source \cite{epl}. Considering the ultra violate cutoff due to lattice,
$T_{source}\left( \mathbf{R}, N \right)$ at $\mathbf{R}=0$ can be approximated by
$\left( 4\pi \right)^{-1}N \log N$ in leading orders in $N$.
Then at $\mathbf{R} = \mathbf{R}_0$, $T_{source}\left( \mathbf{R}-\mathbf{R}_{o}; N_{a} \right)$ is approximately equal to
$(4\pi)^{-1}N_{a}\log N_{a}$ whereas $T_{source}\left( \mathbf{R}_0; N \right) \approx N \phi_{source}( \mathbf{r}_0) $,
where $\phi_{source}\left( \mathbf{r} \right)$ is the potential function for the problem without a sink. Then 
from equation (\ref{lcomb}) we have
\begin{equation}
\frac{1}{4\pi}N_a \log N_a \simeq N \phi_{source}\left( \mathbf{r}_{o} \right). 
\end{equation}
For large $N$, this implies that  $N_a \simeq 4\pi\phi_{source}\left( \mathbf{r}_{o} \right)N/\log N$. In numerical
measurement it is found that for a change of $N$ from $224000$ to $896000$, $N_{a}\log{N}/N$
changes by less than $7\%$ which is consistent with the above scaling relation. For large $N$,
for a sink at a fixed reduced coordinate $\mathbf{r}_o$, the relative size of the defect produced by the
sink decreases as $( \log N)^{-1/2}$. Hence asymptotically, the fractional area of the defect
region will decrease to zero, if sink position $\mathbf{r}_o$ is in a dense patch.

When the sink site is inside a light patch, the subtraction procedure of
equation (\ref{zsink}) gives positive heights, and no longer gives the correct solution.  
However it is observed for patches in the outer layer
where patches are large, effect of sink site is confined within neighboring 
dense patches (Fig.\ref{fig:pstwo}) and rest of the pattern in the asymptotic 
limit remains unaffected.

The pattern  where the source and sink sites are adjacent to each other appears to be
very similar to the one produced in absence of sink site. This is easy to see. 
The Poisson equation analogous to equation (\ref{poisson4}) for this problem is
\begin{equation}
\nabla^2\phi(\mathbf{r})=\Delta\rho(\mathbf{r})-\frac{N}{\Lambda^2}\delta(\mathbf{r})+\frac{N_{a}}{\Lambda^2}\delta(\mathbf{r}-\mathbf{r}_o),
\label{poisson3}
\end{equation}
where $N_a$ is the number of grains absorbed in the sink site at $\mathbf{r}_{o}$. 
In an electrostatic analogy, as discussed earlier, $\phi$ can be considered
as the potential due to a distributed charge of density $-\Delta\rho\left( \mathbf{r}_{o} \right)$
and two point charges of strength $N/\Lambda^{2}$ and $-N_{a}/\Lambda^{2}$ placed
at origin and at $\mathbf{r}_{o}$ respectively.
It is easy to see  that the dominant contribution in the potential is the  monopole term with net
charge $(N- N_a)/\Lambda^2$. The contribution due to other terms decreases as
$1/\Lambda$ for large $\Lambda$, and the asymptotic pattern is the same as without a sink, with $N-N_a$ particles added.

\begin{figure}
\begin{center}
\includegraphics[scale=0.78]{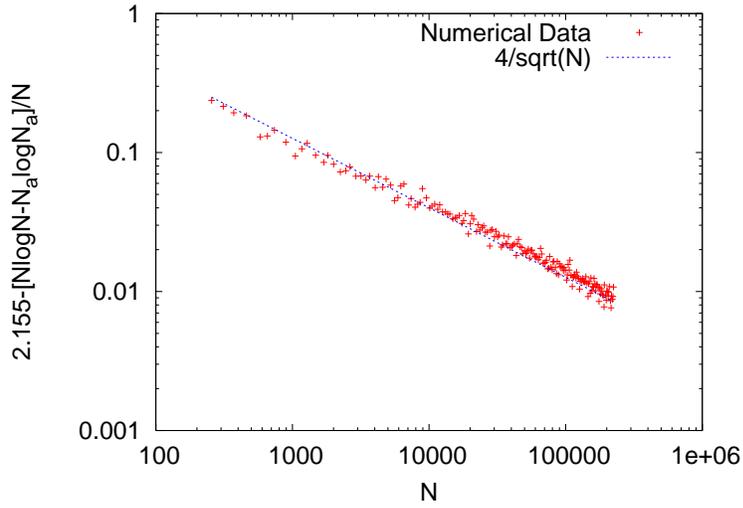}
\end{center}
\caption{Dependence of number of absorbed grains $N_{a}$ on number of grains added $N$ at the origin with
a sink site at $\left( 1, 0 \right)$.}
\label{fig:nan}
\end{figure}

The number of particles absorbed  $N_a$ is determined by the condition that the number
of toppling at $(1,0)$ ( the sink position) is zero. The potential produced by the
areal charge density at $(1,0)$ and $(0,0)$ is nearly the same. The number of toppling
at $(1,0)$ if we add $N_a$ particles at the sink site is approximately $\left (4 \pi\right )^{-1} N_a \log N_a$.
Now, from the solution of the discrete Laplacian, the number of toppling produced at
$(1,0)$ due to $N$ particles added at $(0,0)$ is approximately $\left( 4\pi \right)^{-1}\left (N \log N -C N\right )$ with
$C$ being an undetermined constant. Equating these two, we get
\begin{equation}
 N_a \log N_a  \doteqdot   N \log N - C N
\end{equation}
The above relation is verified with numerical data in figure \ref{fig:nan}. We find that
$(N\log N-N_{a}\log N_{a})/N$ asymptotically approaches a value $C=2.155$ with the
difference from the asymptotic value decreasing as $N^{-1/2}$. As the asymptotic pattern
is the same as produced by adding $\left( N-N_{a} \right)$ grains at the origin without a sink,
we have $N - N_a \simeq \Lambda^2$. Then, using the numerical value for $C$ we get
\begin{equation}
(N -\Lambda^{2})\log ( N -\Lambda^2) \doteqdot  N \log N - 2.155 N.
\label{scaleone}
\end{equation} 
Simplification of this equation for large $N$, shows that $\Lambda$ grows as $\sqrt{N/\log N}$ with $N$. 

\begin{figure}
\begin{center}
\includegraphics[scale=0.75]{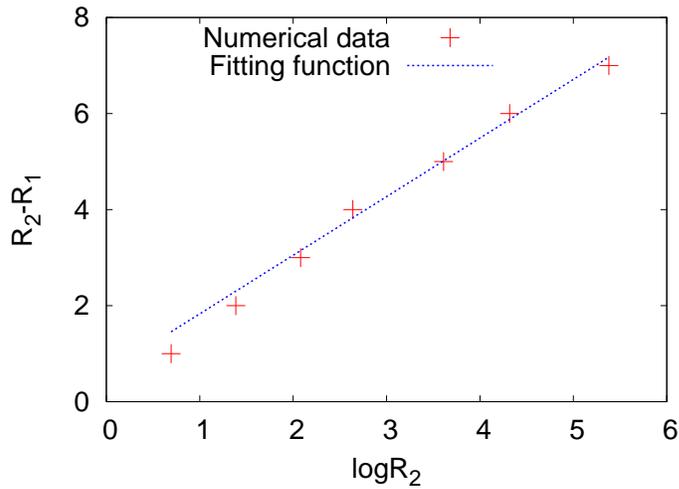}
\end{center}
\caption{Bilateral asymmetry due to the presence of a sink site in Fig.\ref{fig:psone}.}
\label{fig:asymmetry}
\end{figure}

For finite $\Lambda$, the leading  correction to $\phi\left( \mathbf{r} \right)$ comes from
the dipole term in the potential. Presence of this term breaks the reflection 
symmetry of the pattern about the origin. The relative contribution of the dipole potential compared to the monopole
term decays as $\log \Lambda/\Lambda$.  
A measure of the bilateral asymmetry is the
difference of boundary distances on two opposite sides of the source. This 
difference is plotted in Fig.\ref{fig:asymmetry}, where $R_{1}$ and $R_{2}$ are 
boundary distances measured along the positive and negative $x$ axis with a 
sink site placed at $\left( 1,0 \right)$. The difference
$(R_{2}-R_{1})$ is found to fit to   $1.22\log{(R_{2}+0.5)}$. 
\section{Characterization of the pattern with line sink}
We start by recalling the characterization of single source pattern \cite{epl}.
As discussed in Section $3$, the asymptotic patterns can be characterized by the
function $\Delta\rho\left( \mathbf{r} \right)$ in rescaled coordinate.
The single source pattern on F-lattice with chequerboard background is made
of union of distinct regions, called ``patches'', where inside each
patch $\Delta\rho\left( \mathbf{r} \right)$ is constant and takes
only two possible values, $1/2$ in a dense patch and $0$ in a light patch \cite{epl}.

The potential function $\phi\left( \mathbf{r} \right)$ for the single source problem follows the Poisson equation 
\begin{equation}
\nabla^{2}\phi(\mathbf{r})=\Delta\rho(\mathbf{r})-\frac{N}{\Lambda^{2}}\delta(\mathbf{r}).
\label{poisson1}
\end{equation}
The condition that determines $\phi\left( \mathbf{r} \right)$ is the requirement that
inside each patch of constant density, it is a quadratic function 
of $\xi$ and $\eta$ \cite{epl}. Let us write
\begin{equation}
\phi \left( \mathbf{r} \right) = a \xi^{2}+2 h \xi \eta+b \eta^{2}+d\xi+e\eta+f
\label{quadone}
\end{equation}
where $a$, $h$, $b$, $d$, $e$ and $f$ are constants inside a patch and $a+b=\Delta\rho/2$ 
corresponding to the patch. Then each patch is characterized by these parameters. 
Continuity of $\phi\left( \mathbf{r} \right)$ and its derivatives along the boundary between two adjacent
patches imposes linear relations among the corresponding parameters. These linear
equations can be solved on the connectivity graph of patches which forms a square
lattice on two sheeted Riemann surface \cite{epl}.

The pattern with line sink (Fig.\ref{fig:lsone}) retained two important properties present in 
the single source pattern. These are: the asymptotic pattern is made of
union of two types of patches of excess density $1/2$ and $0$
and the separating boundaries of patches are straight lines of slope $0$, $\pm1$ or
$\infty$. However the adjacency graph is changed significantly and this changes the sizes of patches as well. In this section we show how to explicitly determine the potential
function on this adjacency graph.

\begin{figure}
\begin{center}
\includegraphics[scale=0.40]{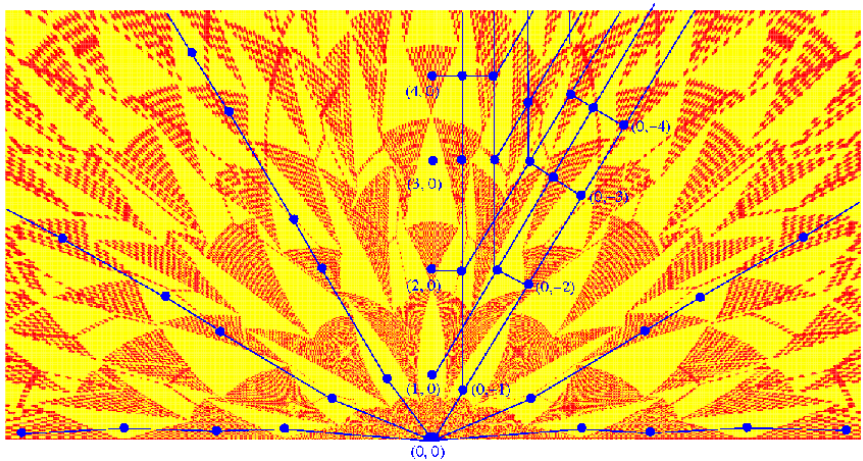}
\caption{$1/r^{3}$ transformation of the pattern in Fig.\ref{fig:lsone}. Two adjoining patches are connected by drawing a straight line.}
\label{fig:1byrsqr}
\end{center}
\end{figure}

\begin{figure}
\begin{center}
\includegraphics[scale=0.28]{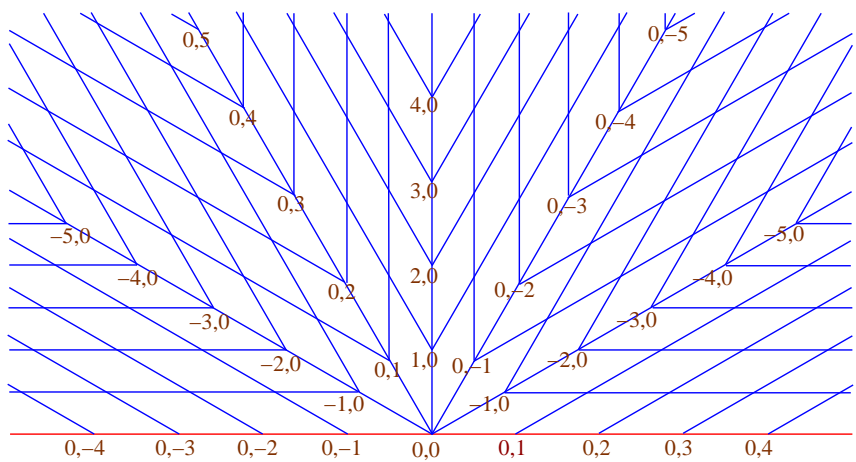}
\caption{Adjacency graph of the patches in pattern in Fig.\ref{fig:lsone}.}
\label{fig:adjLS}
\end{center}
\end{figure}

The adjacency graph of the patches is given in Fig.\ref{fig:adjLS}. 
This representation of the graph is easier to see by taking $1/r^{3}$
transformation of the pattern and then joining the neighboring patches by straight lines (Fig.\ref{fig:1byrsqr}).
Each vertex in the graph is connected to four neighbors except the vertices corresponding to
the patches next to the absorbing line. These have coordination number $3$. Also the vertex
at the center corresponding to the exterior of the pattern is connected to seven neighbors.

Let us write the quadratic potential function in a patch $P$ having excess density $1/2$ as 
\begin{equation}
\phi_{_P}(\mathbf{r})=\frac{1}{8}(m_{_P}+1)\xi^2+\frac{1}{4}n_{_P}\xi\eta+\frac{1}{8}(1-m_{_P})\eta^2 + d_{_P}\xi+e_{_P}\eta+f_{_P},
\label{fform1}
\end{equation}
where the parameters $m$, $n$, $d$, $e$ and $f$ take 
constant values within a patch. Similarly for the lighter patches $P'$
\begin{equation}
\phi_{_{P'}}(\mathbf{r})=\frac{1}{8}m_{_{P'}}(\xi^{2}-\eta^2)+\frac{1}{4}n_{_{P'}}\xi\eta+d_{_{P'}}\xi+e_{_{P'}}\eta+f_{_{P'}}.
\label{fform2}
\end{equation}

Using the continuity of $\phi\left( \mathbf{r} \right)$ and its derivatives along the common boundaries
between neighboring patches it has been shown that for single source pattern
without sink sites $m$ and $n$ take integer values \cite{epl}. Same arguments also applies for this
problem and $\left( m, n \right)$ are the coordinates of the patches in the
adjacency graph in Fig.\ref{fig:adjLS}. These coordinates are shown next to some of the vertices.

There are two different patches corresponding to same set of $\left( m, n \right)$
values. Infact, as in the single source pattern the adjacency graph forms a square lattice
on two sheeted Riemann surface, the same is formed for this pattern but on a
three sheeted Riemann surface. The pattern covers half of the surface with
$\left( m, n \right)$ being the Cartesian coordinates on the surface.

Define function $D\left( m, n \right)=d\left( m, n \right)+ie\left( m, n \right)$ on this
lattice. Using the matching conditions along the common boundaries between neighboring patches it can be
shown that $d$ and $e$ satisfy discrete Cauchy-Riemann condition \cite{epl}
\begin{eqnarray}
d\left( m+1, n+1 \right) - d\left( m, n \right)&=&e\left( m, n+1 \right) - e\left( m+1, n \right), \nonumber \\
e\left( m+1, n+1 \right) - e\left( m, n \right)&=&d\left( m+1, n \right) - d\left( m, n+1 \right),
\label{cr}
\end{eqnarray}
and then the function $D$ follows discrete Laplaces equation
\begin{equation}
\sum_{i=\pm1}\sum_{j=\pm1}D(m+i,n+j)-4D(m,n)=0,
\label{laplace}
\end{equation} 
on this adjacency graph. Let us define $M=m+in$ and $z=\xi+\eta$. Then,
as argued before,  close to the origin the potential $\phi$ diverges as
$1/r$ (equation (\ref{dipole1})). Then the corresponding complex potential
$\Phi\left( z \right) \sim 1/z$. As $M \sim {d^2 \Phi}/{dz^2} $, and
$D \sim d \Phi /dz$,  it follows that for large $|m|+|n|$,
\begin{equation}
D \sim M^{2/3}.
\label{asdls}
\end{equation}
The condition that on the absorbing line $\phi\left( \mathbf{r} \right)$ must vanish implies that for the vertices
with even $n$ along the red line in Fig.\ref{fig:adjLS} $e(0,n)$ vanishes. These
vertices correspond to the patches with absorbing line as horizontal boundary in Fig.\ref{fig:lsone}.
Equation (\ref{laplace}) with above constraint and the boundary condition in equation(\ref{asdls})
has a unique solution. The normalization of $\phi$ is fixed by the requirement that
$d(1,0)=-1$ which fixes the diameter of the pattern to be $2$ in reduced units.

Equation \ref{laplace} is the standard two-dimensional lattice Laplace equation,
whose solution is well-known when $\left( m, n \right)\in\mathbb{Z}^{2}$ \cite{spitzer}.
In our case when the lattice sites form a surface of two sheets, we have not been able to find a 
closed-form formula for $D(m,n)$. However the solution can be determined 
numerically to very good precision by solving it on a finite grid $-L\le m,n\le L$
with the above conditions imposed exactly at the boundary. The
calculation is performed with $D=M^{2/3}$ at the boundary and then the solution is normalized
to have $d\left( 1, 0 \right)=-1$.
We determined $d$ and $e$ numerically for $L=100$, $200$, $300$, $400$ and $500$ and
extrapolated our results for $L\rightarrow \infty$. 

\begin{figure}
\begin{center}
\includegraphics[scale=0.4]{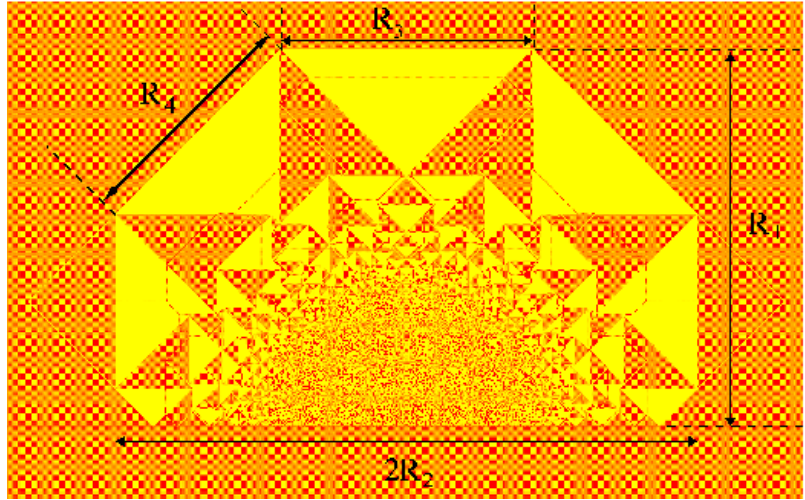}
\end{center}
\caption{Spatial lengths $R_{1}$, $R_{2}$, $R_{3}$ and $R_{4}$ tabulated in Table $1$.}
\label{fig:LSlengths}
\end{figure}

Comparison of results from this numerical calculation and that obtained by
measurements on the pattern is presented in Table $1$. 
We considered four different lengths $R_{1}$, $R_{2}$, $R_{3}$ and $R_{4}$ on the
pattern and they are shown in Fig.\ref{fig:LSlengths}. Among them, according to the
definition of the diameter of the pattern, $R_{1}=2\Lambda$. We present values
of $R_{2}$, $R_{3}$ and $R_{4}$ normalized by $R_{1}$ for different $N$. Theoretical values of these lengths
are determined from asymptotic values of $d$ and $e$. Comparision of these results
shows very good agreement among the theoretical and measured values.
%%%%%%%%%%%%%%%%%%%%%%%%%%%%%%%%%%%%%%%%%%%%%%%%%%%%%%%%%%%%%%%%%%
\begin{table}
  \begin{center}
    \begin{tabular}{|c||c|c|c|c|c|}
      \hline
      ~N~ & $896k$ & $14336k$ & $57344k$ & $229376k$ & Theoretical \\
      \hline
      \hline
      $\frac{R_{2}}{R_{1}}$ & 0.769 & 0.768 & 0.770 & 0.770 & 0.7698  \\
      \hline
      $\frac{R_{3}}{R_{1}}$ & 0.675 & 0.675 & 0.667 & 0.668 & 0.6666 \\
      \hline
      $\frac{R_{4}}{R_{1}}$ & 0.609 & 0.609 & 0.617 & 0.616 & 0.6172 \\
      \hline
    \end{tabular}
    \caption{Comparison of different lengths measured directly from the pattern in Fig.\ref{fig:LSlengths}
for increasing values of $N$, with their theoretical values.}
  \end{center}
  \label{table:first}
\end{table}
%%%%%%%%%%%%%%%%%%%%%%%%%%%%%%%%%%%%%%%%%%%%%%%%%%%%%%%%%%%%%%%%%%%%%%%%%%%
\section{Patterns with two sources}
In this section we discuss patterns produced by adding $N$ grains each at two
sites placed at a distance $2\Lambda\mathbf{r_o}$ from each other along the $x$-axis
at $\Lambda\mathbf{r}_{o}$ and $-\Lambda\mathbf{r}_{o}$ with $\mathbf{r}_{o}\equiv \left( \xi_{o}, 0 \right)$. 
Again the diameter $2\Lambda$ is defined as the height of the smallest
rectangle enclosing all sites that have toppled atleast once.
Two limits $r_0$ close to zero and $r_0$ large are trivial:
For $r_o\rightarrow 0$, the asymptotic pattern is same as that produced by 
adding grains at a single site. On the otherhand if $r_{o} > 1$,  each source produces its own pattern, which do not overlap, 
and the final pattern is simple superposition of the two patterns. 
\begin{figure}
\begin{center}
\includegraphics[scale=0.23]{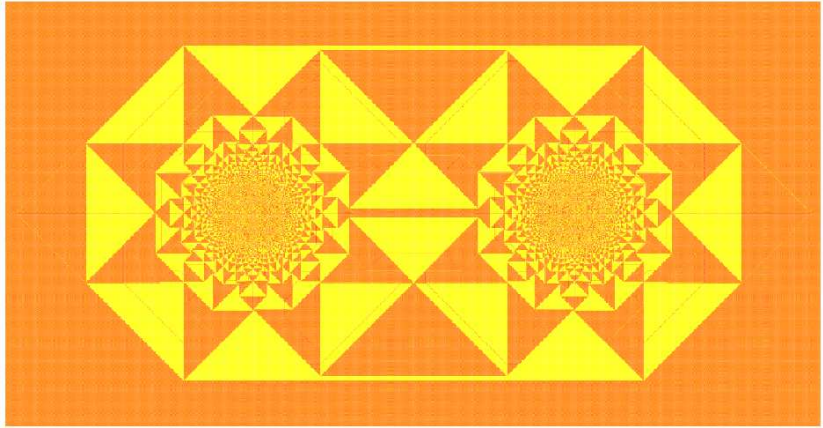}
\caption{Pattern produced by adding $N=640000$ grains each at ($-760,0$) and 
($760,0$) on F-lattice with initial chequerboard distribution of grains and 
relaxing. This corresponds to $r_{o}=0.95$. Color code red=0 and yellow=1.
(Details can be seen in the online version using zoom in )}
\label{fig:twosource}
\end{center}
\end{figure}
\begin{figure}
\begin{center}
\includegraphics[scale=0.14]{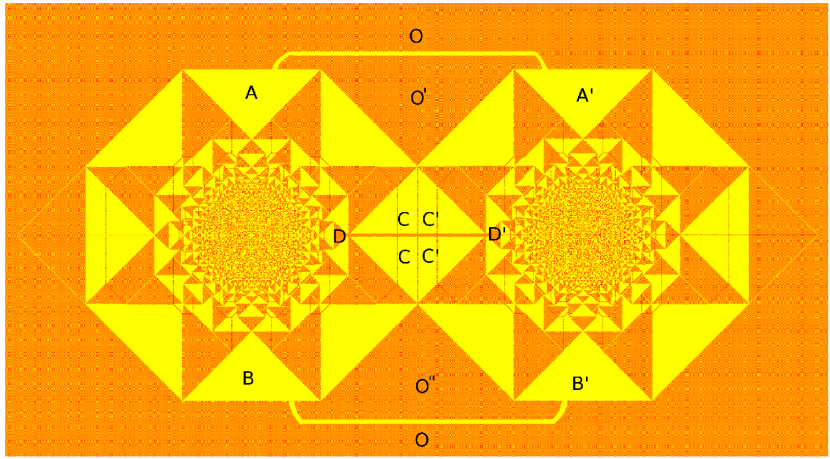}
\caption{Pattern constructed by combining two single source patterns and drawing
connecting lines between few patches following the connectivity in the pattern in 
Fig.\ref{fig:twosource}.}
\label{fig:edited}
\end{center}
\end{figure}

\begin{figure}
\begin{center}
\includegraphics[scale=0.13]{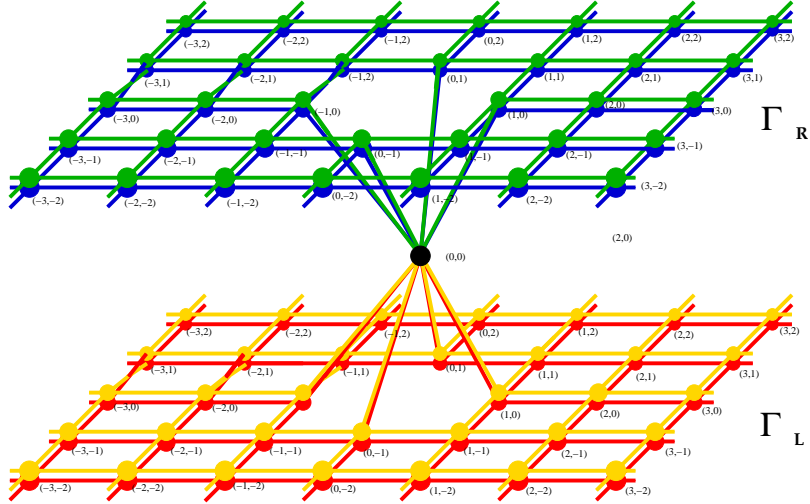}
\end{center}
\caption{Representation of adjacency graph of patches for two non-overlapping single source patterns
as square grid  on two Riemann surfaces each of two sheets. Vertices with the same $(m,n)$ coordinates on different sheets are represented by different colors.}
\label{fig:adjnocol}
\end{figure}

As noted before, the connectivity graph for single source
pattern has square lattice structure on a Riemann surface of two sheets \cite{epl}. Then
the graph for two non-intersecting single source pattern is square 
lattice on two disjoint Riemann surfaces each consists of two sheets 
(Fig.\ref{fig:adjnocol}). Only the vertex at the origin
represent the exterior of the pattern, which is same for both the
single source patterns. It has sixteen neighbors and is placed
midway between the two Riemann surfaces.
For later convenience let us associate the lower Riemann surface to the pattern
around $-\mathbf{r}_o$ and denote it by $\Gamma_L$. Similarly the
upper Riemann surface as $\Gamma_R$ corresponding to the pattern around
$\mathbf{r}_o$.

For $ 0 < r_{o} <1$, using the Abelian property, we can first topple 
as if the other source was absent. The resulting pattern still has some
unstable sites in the region where the patterns overlap. 
Further relaxing these sites transfers these excess grains outward,
and changes the dimensions and positions of patches: some patches become bigger,
some may merge, and some times a patch may beak into two disjoint patches.  

The pattern produced with two sources with $r_0 = 0.95$ is shown in Fig.\ref{fig:twosource}. We see that there are still
only two types of periodic patches, corresponding to $\Delta\rho(\mathbf{r})$ values $0$
and $1/2$,  and the slopes of the boundaries between patches takes values 
$0$, $\pm1$ or $\infty$. 

\begin{figure}
\begin{center}
\includegraphics[scale=0.13]{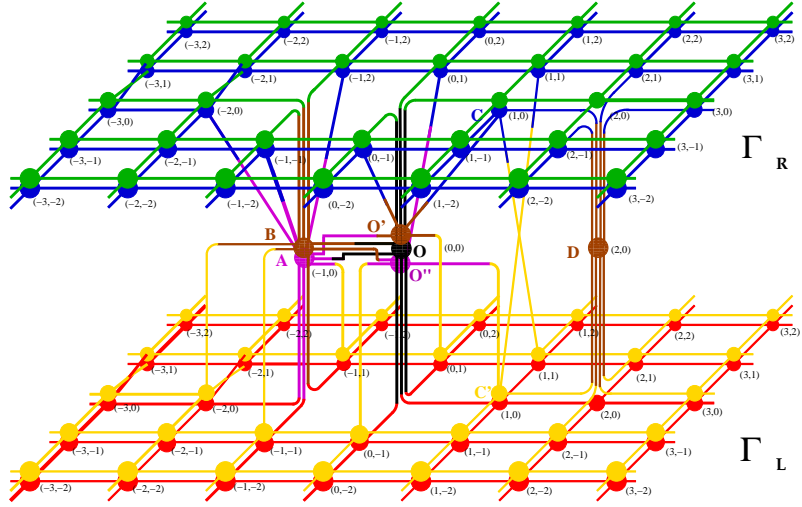}
\end{center}
\caption{Connectivity graph for two intersecting single source patterns around two 
sites of addition placed at a distance $2 \mathbf{r}_o$ from each other. The graph
has the structure of square grids on four Riemann sheets except for a finite number
of vertices indicated by alphabates $A$, $B$, $O$, $O'$, $O"$ and $D$ shown placed
in middle layer. This graph remains unchanged for $r_{o}$ in the range $0.65$ to $1.00$.}
\label{fig:adj}
\end{figure}

The relaxations due to overlaps change the adjacency graph  from the case with 
no  overlap. This modified adjacency graph is shown in Fig.\ref{fig:adj}.
However, for $r_0$ just below $1$, these changes are few, and are listed below.

i) We note that patches labelled $A$ and $A'$ in Fig.\ref{fig:edited} have the
same potential function $\phi$. Then, for $r_0$ just below $1$, these patterns can
join with each other by a thin  strip. This only requires a small movement in the
boundaries of nearby patches, ( i.e.  only a small change  in the $d$ and $e$ values
of nearby patches). Thus, in the adjacency graph, the vertices corresponding to $A$
and $A'$ are collapsed into a single vertex $A$ in Fig.\ref{fig:adj}. 

ii) Similarly, the vertices corresponding to patches $B$ and $B'$ in 
Fig.\ref{fig:edited} are collapsed into a single vertex $B$ in Fig.\ref{fig:adj}.

iii) This divides the region outside the pattern in three parts, $O$, $O'$ and $O"$.
They are also shown in Fig.\ref{fig:adj} as separate vertices.

iv) The patches marked $C$ and $C'$ also have the same quadratic form, and the
vertical boundary between them disappears. However, the patches $D$ and $D'$ are
also joined by a thin strip. This horizontal strip divides the joined $C$ and $C'$
into two again (Fig.\ref{fig:edited}). 

\begin{figure}
\begin{center}
\includegraphics[scale=0.20]{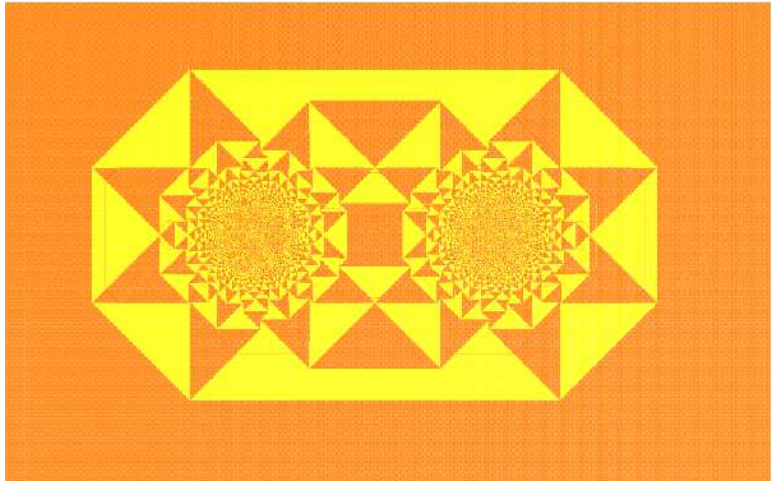}
\end{center}
\caption{Pattern produced by adding $640000$ grains at site $\left( -600,0 \right)$
and $\left( 600,0 \right)$. Although the pattern is significantly different from
the one in Fig.\ref{fig:twosource}, their adjacency graph is same.}
\label{fig:bellow}
\end{figure}

The adjacency of other patches remains unchanged. The adjacency graph of the
pattern is shown in Fig.\ref{fig:adj}. Interestingly, this new adjacency graph
remains the same for all $0.70 < r_0 < 1$, even though  for $r_{o} < 0.85$,
the sizes of different patches are substantially different. [Compare the pattern
for $r_0 =0.70$ in Fig.\ref{fig:bellow}, with pattern for $r_0=0.95$ in Fig.\ref{fig:twosource}].
Shape of the patches near the center of Fig.\ref{fig:bellow} is different
from that in Fig.\ref{fig:twosource}.

In Fig.\ref{fig:adj}, we have have placed  the vertices which are formed by merging
or dividing patches in the overlap region  midway between the Riemann sheets corresponding
to the two sources. As $r_0$ is  decreased  below $0.70$, more collisions between
growing patches will occur, and the number of vertices in this middle region will increase.
For any nonzero $r_0$, the number of vertices in the middle layer is finite. In the
$r_{o}\rightarrow 0$ limit vertices from both the surfaces $\Gamma_{L}$ and $\Gamma_{R}$
come together and form a single Riemann surface corresponding to a single source pattern
around $\mathbf{r}=0$. For $r_0$ small, but greater than zero, the outer patches are arranged
as in the single-source case, but closer to the sources, one has a crowded pattern near each
source. In the adjacency graph, this corresponds to vertices near the patch $(0,0)$ roughly
arranged as on a Riemann surface of two sheets, while the ones farther from the patch $(0,0)$
remain undisturbed on the 4-sheeted Riemann surface.

We now characterize the pattern  with two sources, and $r_0 > 0.70$ in detail by explicitly determining
the potential function on this  adjacency graph.

The Poisson equation analogous to Eq.(\ref{poisson1}) for this problem is
\begin{equation}
\nabla^2\phi(\mathbf{r})=\Delta\rho(\mathbf{r})-\frac{N}{\Lambda^2}\delta(\mathbf{r}-\mathbf{r}_o)-\frac{N}{\Lambda^2}\delta(\mathbf{r}+\mathbf{r}_o).
\label{poisson2}
\end{equation}
Let us use the same quadratic form of the potential
function given in equation (\ref{fform1}) and equation (\ref{fform2}).

Again using the same argument given in \cite{epl} it can be shown that
$m$ and $n$ are the coordinates of the patches in both the adjacency graphs
in Fig.\ref{fig:adjnocol} and Fig.\ref{fig:adj}. These coordinates are shown  next
to each vertex. Also, in the region away from the origin, on each sheet,the function $D(m,n)=d(m,n)+ie(m,n)$  satisfies the discrete Laplace equation
\begin{equation}
\sum_{i=\pm1}\sum_{j=\pm1}D(m+i,n+j)-4D(m,n)=0.
\label{laplace2}
\end{equation} 
Let us define $z_o=\xi_o+i\eta_o$ where $(\xi_o,\eta_o)$ and
$(-\xi_{o},-\eta_{o})$ are the coordinates corresponding to $\mathbf{r}_{o}$ and $-\mathbf{r}_{o}$.
Considering that close to $\mathbf{r}_{o}$ and $-\mathbf{r}_{o}$ the potential
$\phi\left( \mathbf{r} \right)$ diverges logarithmically it can be shown (as done for single source pattern
in \cite{epl}) that for large $|m|+|n|$,
\begin{eqnarray}
D(m,n)&=&\pm\frac{A}{\sqrt{2\pi}}\sqrt{M}+ \bar{z}_o\frac{M}{4}\rm{, ~on~ }\Gamma_L\nonumber \\
&=&\pm\frac{A}{\sqrt{2\pi}}\sqrt{M}- \bar{z}_o\frac{M}{4}\rm{, ~on~  }\Gamma_R
\label{asymptot}
\end{eqnarray}
where $A$ is a constant independent of $N$ or $\Lambda$. 

Again we determine the solution of equation (\ref{laplace2})
numerically to very good precision by solving it on finite grid $-L\le m,n\le L$
with the conditions Eq. (\ref{asymptot})
imposed exactly at the boundary. The value of $A$ is determined from a self 
consistency condition that the diameter of the pattern in reduced
coordinate is 2 which imposes $2e(-1,0)=-1$ corresponding to the vertex $A$
in Fig.\ref{fig:adj}. We determined $d$ and $e$ numerically
for $L=100$, $200$, $300$, $400$ and $500$ and extrapolated our results for 
$L\rightarrow \infty$. 

\begin{figure}
\begin{center}
\includegraphics[scale=0.3]{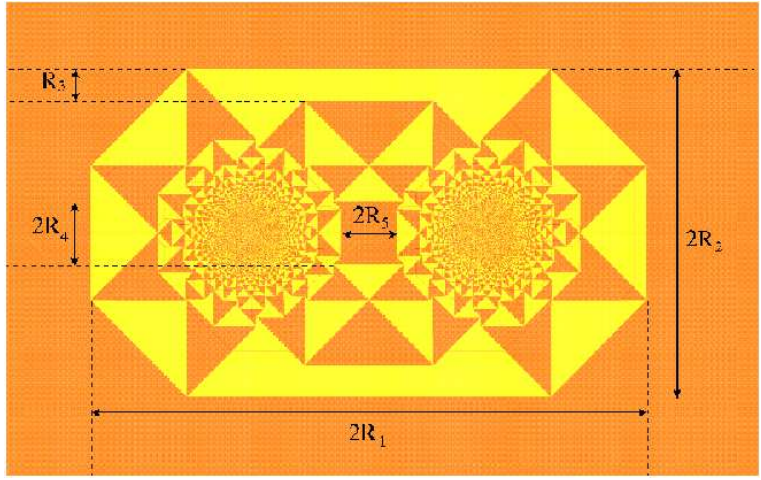}
\end{center}
\caption{Spatial lengths $R_{1}$, $R_{2}$, $R_{3}$, $R_{4}$ and $R_{5}$ tabulated in Table.$2$}
\label{fig:lengths}
\end{figure}

Comparison of results from this numerical calculation and that obtained by
measurements on the pattern in presented in Table $2$. We
considered five different lengths in the pattern corresponding to $r_{o}=0.800$.
These different lengths are drawn in Fig.\ref{fig:lengths} and their values rescaled by $\sqrt{N}$, for the patterns
with increasing $N$, is given in Table $2$. Theoretical results are
obtained using the asymptotic values of $d$ and $e$ for large $L$. The rescaled
lengths extrapolated to infinite $N$ limit matches very well to the theoretical results.
%%%%%%%%%%%%%%%%%%%%%%%%%%%%%%%%%%%%%%%%%%%%%%%%%%%%%%%%%%%%%%%%%%
\begin{table}
  \begin{center}
    \begin{tabular}{|c||c|c|c|c|c|c|}
      \hline
      ~N~ & $2.5k$ & $10k$ & $40k$ & $160k$ & $640k$ & Theoretical \\
      \hline
      \hline
      $\frac{R_{1}}{\sqrt{N}}$ & 1.84 & 1.84 & 1.84 & 1.83 &  1.83 & 1.82  \\
      \hline
      $\frac{R_{2}}{\sqrt{N}}$ & 1.06 & 1.07 & 1.07 & 1.06 & 1.05 & 1.06 \\
      \hline
      $\frac{R_{3}}{\sqrt{N}}$ & 0.22 & 0.21 & 0.20 & 0.19 & 0.18 & 0.18 \\
      \hline
      $\frac{R_{4}}{\sqrt{N}}$ & 0.18 & 0.19 & 0.19 &  0.18 & 0.18 & 0.18 \\
      \hline
      $\frac{R_{5}}{\sqrt{N}}$ & 0.20 & 0.22 & 0.21 & 0.21 & 0.21 & 0.21 \\
      \hline
    \end{tabular}
    \caption{Comparison of different lengths measured directly from the two source pattern
for $r_{o}=0.800$ with their theoretical values.}
  \end{center}
  \label{table:second}
\end{table}
%%%%%%%%%%%%%%%%%%%%%%%%%%%%%%%%%%%%%%%%%%%%%%%%%%%%%%%%%%%%%%%%%%%%%%%%%%%
\section{Summary}
We have shown that exact characterization of the patterns
in F-lattice on chequerboard background reduces to solving  a discrete
Laplace equation on the adjacency  graph of the pattern. For the single source pattern
this graph is a square grid on two-sheeted Riemann surface and in
presence of a line sink it is on a $3$-sheeted Riemann surface.
This Riemann surface structure occurs for other sink geometries also
and the number of sheets can be determined from the way
$\phi$ diverges near origin. 

If the potential $\phi(r)$ diverges as $ r^{-a}$ near the origin, then the corresponding
complex function $\Phi(z) \sim z^{-a}$.  Then $\frac {d^2}{dz^2} \Phi \sim z^{-2-a}$. In
all the cases we studied above, the patch to which point $z$ belongs is characterized by
integers $(m,n)$, where $\frac {d^2}{dz^2} \Phi  \sim m+in$. Also $\frac{d}{dz} \Phi \sim
d+ie $. Writing $D=d+i e$, and $M = m+in$, we see that $D \sim M ^{\frac{1+a}{2+a} }$.
This then gives the number of Riemann sheets. For example, for a wedge angle
$\omega=2\pi$, we have $a=1/2$. Then $D \sim M^{3/5}$, and the Riemann surface would have 5 sheets. 
\begin{figure}
\begin{center}
\includegraphics[scale=0.21]{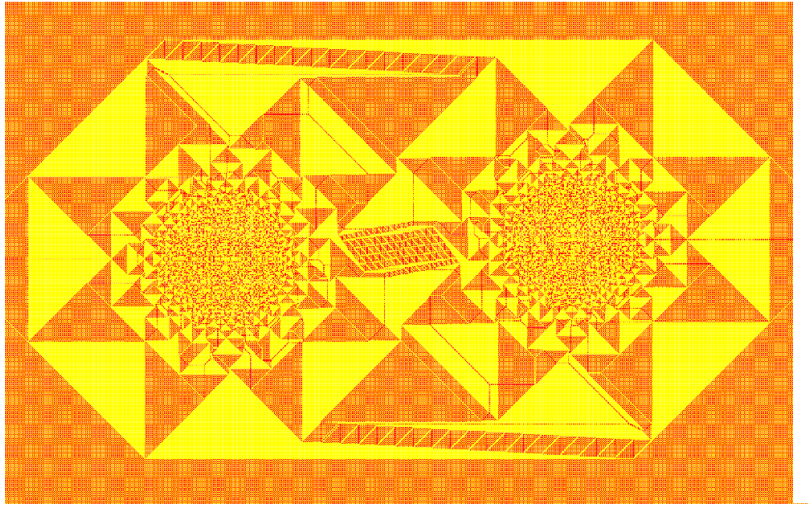}
\caption{Pattern produced by adding $N=40000$ grains each at ($-180,0$) and 
($180,20$) on F-lattice with initial chequerboard distribution of grains and 
relaxing. Color code red=0 and yellow=1. (Details can be seen in the online
version using zoom in )}
\label{fig:distr}
\end{center}
\end{figure}

The cases where the full pattern can be explicitly determined are clearly special. 
For example, one of the conditions used for exact characterization of patterns in this paper is that
inside each patch the height variables are periodic and hence $\Delta\rho\left( \mathbf{r} \right)$
is constant. It is easy to check that this condition is not met for most of the sink geometries.
For example, patterns  of the type discussed in Section $4$
with any $\omega$ other than integer multiples of $\pi/4$ has aperiodic
patches. Similarly for the patterns with two sources even slight deviation
of the position of the second source in Fig.\ref{fig:twosource} from the x-axis
introduces aperiodic patches. One of such patterns produced by adding $40000$
grains each at $\left( -180, 0 \right)$ and $\left( 180, 20 \right)$ is shown
in Fig.\ref{fig:distr}. The regions with stipes of red and yellow are the
aperiodic patches. Also boundary of all these aperiodic patches
have slopes other than $0$, $\pm 1$ and $\infty$ and some  boundaries of patches are not  straight lines.  In such cases, the present treatment is clearly not applicable.  However the scaling analysis for the growth of
spatial lengths in the pattern with $N$ is still valid.

The function $D=d+ie$ satisfies discrete Cauchy-Riemann condition (equation (\ref{cr})). These functions are known as discrete  holomorphic
functions in the mathematics literature. They have been  usually  studied  for a square grid of 
points on the plane \cite{duffin,spitzer}. While more general discretizations of the plane have been discussed  \cite{mercat,laszlo}, not much is known about the behavior of such functions  for muti-sheeted Riemann surfaces.

In our analysis we have also used a fact that the patterns have non-zero average overall 
excess density ( i.e. $C_2$ in Eq. (8) is nonzero). 
The case  when $C_2$ is zero is quite different, and requires a substantially different treatment.
We hope to discuss such patterns in a future publication \cite{next}.

\begin{acknowledgements}
DD thanks Prof. A. Libchaber for suggesting this problem, and some useful discussions, and the Department of Science and Technology, India for financial support through a J.C. Bose fellowship.
\end{acknowledgements}

\end{document}